\documentstyle{article}
\title{
\begin{flushright}
{\bf\normalsize   COLO-HEP-269}\\
\end{flushright}
\bf A Numerical Test of KPZ Scaling:\\
Potts Models Coupled to Two-Dimensional Quantum Gravity
}

\author{ {\it C.F. Baillie} \\
         Physics Dept. \\
         University of Colorado\\
         Boulder, CO 80309, USA\\
	 \\
         and \\
	 \\
         {\it D.A. Johnston}\\
         Dept. of Mathematics\\
         Heriot-Watt University\\
         Riccarton\\
         Edinburgh, EH14 4AS, Scotland}
\begin{document}
  \maketitle
                      {\Large
                      \begin{abstract}
%
We perform Monte Carlo simulations using the Wolff cluster algorithm
of the q=2 (Ising), 3, 4 and q=10 Potts models on
dynamical phi-cubed graphs of spherical topology with up to 5000 nodes.
We find that the measured critical exponents
are in reasonable agreement with those from the exact solution of the Ising
model and with those
calculated from KPZ scaling for q=3,4 where no exact solution is available.
Using Binder's cumulant we find that the q=10 Potts model
displays a first order phase transition on a dynamical graph, as it does on a
fixed lattice.
We also examine the internal
geometry of the graphs generated in the simulation,
finding a linear relationship between ring length probabilities and the
central charge of the Potts model.
\\
\\
To appear in Modern Physics Letters A.
%
                        \end{abstract} }
%
  \thispagestyle{empty}
%
%
  \newpage
%
                  \pagenumbering{arabic}

\section{Introduction}

There has been considerable activity recently in the field of two-dimensional
matter
coupled to two-dimensional gravity, motivated initially by string theory.
Both the continuum Liouville theory
and matrix models have been used in
these investigations. The work in \cite{1}
by Knizhnik, Polyakov and Zamolodchikov (KPZ) and in \cite{2} by Distler, David
and Kawai (DDK), with
the light-cone and conformal gauge-fixed Liouville theories respectively,
allowed the
calculation of critical exponents for conformal field theories
with central charge $c<1$ coupled to two-dimensional quantum gravity.

Both \cite{1},\cite{2} showed that the effect of coupling such theories to
gravity was to ``dress''
an operator of conformal weight $\Delta_0$
in the original theory without gravity yielding a new weight $\Delta$ given by
\begin{equation}
\Delta - \Delta_0 = - {\alpha^2 \over 2} \Delta ( \Delta - 1),
\label{e0}
\end{equation}
where
\begin{equation}
\alpha = - { 1 \over 2 \sqrt{3} } ( \sqrt{ 25 -c } - \sqrt{ 1 -c} ).
\label{e01}
\end{equation}
{}From eq. \ref{e0}, which is called the KPZ scaling relation, we can see that
the weights
are modified by the gravitational dressing in a manner that depends only on the
central charge. We can thus calculate the weights of operators when coupled to
gravity
by referring to the usual Kac table \cite{3} to get $\Delta_0$ and then using
eq. \ref{e0} to
find $\Delta$. The $q=2,3,4$ Potts models which have $c={1 \over 2},{4 \over
5},1$ respectively
fall within the framework discussed above, with the $q=4$ model lying on the
boundary
of the strong-coupling region $1 < c < 25$ where KPZ scaling breaks down. The
$q=10$ model has a first order transition on a fixed lattice and conformal
field theory
methods are therefore not applicable.

If we denote the critical temperature for a continuous spin-ordering
phase transition by $T_c$ and the reduced temperature $|T - T_c|/T_c$ by $t$
then the
critical exponents $\alpha, \beta, \gamma, \nu, \delta, \eta$
can be defined in the standard manner as $t \rightarrow 0$
\begin{eqnarray}
C \simeq t^{-\alpha} \; &;& \; M \simeq t^{\beta}, T<T_c \nonumber \\
\chi \simeq t^{- \gamma} \; &;& \;
\xi \simeq t^{- \nu} \nonumber \\
M(H,t=0) &\simeq& H^{1 / \delta}, \; H \rightarrow 0 \nonumber \\
<M(x) M(y)> &\simeq& {1 \over |x - y |^{d -2 + \eta}}, \; t=0
\label{e04}
\end{eqnarray}
where $C$ is the specific heat, $M$ is the magnetization,
$\chi$ is the susceptibility, $\xi$ is the correlation length and
$H$ is an external field.
In the theories without gravity
it is possible to calculate $\alpha$ and $\beta$
using the conformal weights of the energy density operator and spin operator
(for a review see \cite{4}).
Given these
we can now use the various scaling relations \cite{3}
\begin{eqnarray}
\alpha &=& 2 - \nu d \nonumber \\
\beta  &=& {\nu \over 2} (d - 2 + \eta) \nonumber \\
\gamma &=& \nu ( 2 - \eta) \nonumber \\
\delta &=& { d + 2 - \eta \over d - 2 + \eta }
\label{e033}
\end{eqnarray}
to obtain the other exponents.

When we couple the conformal field theories to gravity
we can still calculate $\alpha$ and $\beta$ using the new conformal weights
given by
KPZ scaling. Then,
provided the scaling
relations in eq. \ref{e033} are still valid, we can obtain the full set of
exponents.
For reference
we have listed the critical exponents for the $q=2,3,4$ Potts models in
Table 1 and the $q=2,3,4$ Potts models coupled to gravity in Table 2.


\begin{center}
\begin{tabular}{|c|c|c|c|c|c|c|c|} \hline
$q$& $c$          & $\alpha$     & $\beta$       & $\gamma$  & $\delta$& $\nu$
& $\eta$ \\[.05in]
\hline
$2$& $\frac{1}{2}$& $0$          & $\frac{1}{8}$ & $\frac{7}{4}$ & $15$& $1$
      & $\frac{1}{4}$ \\[.05in]
\hline
$3$& $\frac{4}{5}$& $\frac{1}{3}$& $\frac{1}{9}$ & $\frac{13}{9}$& $14$&
$\frac{5}{6}$& $\frac{4}{15}$\\[.05in]
\hline
$4$& $1$          & $\frac{2}{3}$& $\frac{1}{12}$& $\frac{7}{6}$ & $15$&
$\frac{2}{3}$& $\frac{1}{2}$ \\[.05in]
\hline
\end{tabular}
\end{center}
\vspace{.1in}
\centerline{Table 1: Analytical exponents for 2d Potts models.}
\bigskip

\begin{center}
\begin{tabular}{|c|c|c|c|c|c|c|c|} \hline
$q$& $c$          & $\alpha$      & $\beta$      & $\gamma$     & $\delta$&
$\nu$    & $\eta$ \\[.05in]
\hline
$2$& $\frac{1}{2}$& $-1$          & $\frac{1}{2}$& $2$          & $5$&
$\frac{3}{d}$ & $2-\frac{2d}{3}$\\[.05in]
\hline
$3$& $\frac{4}{5}$& $-\frac{1}{2}$& $\frac{1}{2}$& $\frac{3}{2}$& $4$&
$\frac{5}{2d}$& $2-\frac{3d}{5}$\\[.05in]
\hline
$4$& $1$          & $0$           & $\frac{1}{2}$& $1$          & $3$&
$\frac{2}{d}$ & $2-\frac{d}{2}$ \\[.05in]
\hline
\end{tabular}
\end{center}
\vspace{.1in}
\centerline{Table 2: Analytical exponents for Potts models coupled to
2d quantum gravity.}

\bigskip

\noindent
Note that for the latter $d$, the internal fractal dimension of the graph,
is not known a priori so
$\nu$ and $\eta$ are obtained as functions of $d$.
Reassuringly, the standard scaling relations {\it are} satisfied for the one
model that
has been exactly solved when it is coupled to gravity - the Ising model. The
critical
exponents $\alpha$ and $\beta$ calculated from the exact solution
agree with those calculated by KPZ and the full set of exponents satisfy the
relations in eq. \ref{e033}.

The exact solution
of the Ising model coupled to gravity \cite{4a},\cite{5} made
use of the matrix-model techniques pioneered in \cite{6}
by showing that the partition function on a random graph
was equal to the free energy of
a two-hermitean $N \times N$ matrix model.
The matrix model with a cubic interaction generates phi-cubed graphs with two
types of vertices
representing the spins, so the
sum over graphs
is equivalent to integrating over the metric when coupling the spins to
two-dimensional gravity.
The model was solved exactly in the planar limit $N \rightarrow
\infty$ with both a cubic interaction and a quartic interaction. Both
interactions gave
a third order magnetization transition with the critical exponents shown in
Table 2.
It was found that the inverse critical temperature
$\beta_c = {1 \over 2} ln{108 \over 23} = 0.7733185$
for cubic interactions with no tadpoles or self-energies \cite{6a}.
For the Potts models on random graphs with a fixed number of nodes
$N$ the partition function $Z_N$ is
\begin{equation}
Z_N =  \sum_{G^{(N)}} \sum_{\sigma} \exp \left( - {\beta \over 2}
\sum_{i,j=1}^N G^{(N)}_{ij}
\delta( \sigma_i \sigma_j) \right)
\label{e2}
\end{equation}
where
$G^{(N)}$ is the adjacency matrix,
$\delta$ is a Kronecker delta and
$\beta$ is the inverse temperature $1/T$ (not to be confused with the critical
exponent $\beta$!)
We have, in general, $q$ species of spin taking the values $0,1,...,q-1$.
The solution of these models
along the lines of \cite{4a},\cite{5}
has so far proved elusive for $q>2$ \cite{7} so we do
not know the order of the phase transition or the critical temperature for
$q=3,4$.

Simulations of the Ising model have been carried out on both dynamical
triangulations
\cite{8a},\cite{8b} and phi-cubed graphs \cite{9}
(these are effectively a numerical evaluation of eq. \ref{e2})
and satisfactory agreement between the
measured and theoretical values of the critical exponents found.
However, no previous numerical work has been carried out on the $q=3,4$ Potts
models where only the
KPZ results are available rather than the exact solution.
The object of the simulations
in this paper is to measure the critical exponents in
the $q=3,4$ Potts models
on dynamical phi-cubed graphs (i.e. coupled to two-dimensional quantum gravity)
in order to see if they are in agreement with the values calculated
using KPZ scaling.
Our work can thus be considered as a numerical test of the validity of KPZ
scaling
in these models.
We also investigate the $q=10$ Potts model to see if it has a first order
transition
on a dynamical phi-cubed graph.
The Ising model, where we have the exact solution and previous simulations to
compare with,
is used to verify that our simulation is working properly.
We have chosen to simulate phi-cubed graphs rather
than the dual triangulation because the exact solution of the Ising model is
couched in this
form, although universality would lead us to believe that the two should give
identical
results.
The spin model critical exponents are unaffected by the topology of the
phi-cubed graphs so
we use
graphs of spherical topology for simplicity.

\section{Simulations}

We perform a microcanonical (fixed number of nodes)
Monte Carlo simulation on graphs with
$N = 50,100,200,300,500,1000,2000$ and $5000$ nodes,
at various values of $\beta$ between $0.1$ and $1.5$.
The Monte Carlo update consists of two parts:
one for the spin model and one for the graph.
For the Potts model we use a cluster update algorithm
since it suffers from much less critical slowing down than the
standard Metropolis algorithm  (for a review see \cite{10cfb}).
There are two popular cluster algorithm
implementations - Wolff \cite{10a} and Swendsen-Wang \cite{10b}.
As they are equivalent for the usual two-dimensional $q=2,3$ Potts
models \cite{10c} we use Wolff's variant because it is computationally faster.
In order to keep the autocorrelation time, in terms of update sweeps,
constant the number of times the Wolff cluster algorithm is applied per
sweep, $W$, must be increased as the temperature increases (since the
average cluster size decreases).
At the critical temperature, where the correlation length diverges,
the integrated autocorrelation time takes on its maximum value which scales as
\begin{equation}
\tau_{int} \simeq N^{z \over d},
\end{equation}
where $z$ is the dynamical critical exponent.
The measured values of $\tau_{int}$ and fitted values of $z$ are listed in
Table 3.

\begin{center}
\begin{tabular}{|c|c|c|c|} \hline
$N$   & $q=2$    & $q=3$    & $q=4$      \\[.05in]
\hline
500   & 1.38(19) & 3.83(17) & 8.31(5)    \\[.05in]
1000  & 1.45(13) & 4.35(17) & 9.17(5)    \\[.05in]
2000  & 1.51(10) & 4.40(20) & 10.18(16)  \\[.05in]
5000  & 1.58(8)  & 4.73(26) & 9.2(2)$^*$ \\[.05in]
\hline
$z/d$ & 0.06(5)  & 0.09(3)  & 0.15(1)  \\[.05in]
\hline
\end{tabular}
\end{center}
\vspace{.1in}
\centerline{Table 3: $\tau_{int}$ at $\beta_c$ for $N=500,1000,2000,5000$ and
fitted}
\centerline{values of $z/d$ (last line); $^*$ indicates that value is not
reliable}
\centerline{due to $W$ being too large and run time being too short.}
\bigskip

\noindent
If we assume that $d$ is $2$ (or $3$) then we obtain
$z = 0.12(10), 0.18(6), 0.30(2)$ (or $0.18(15), 0.27(9), 0.45(3)$) for
$q=2,3,4$ Potts models, respectively.
(This can be compared with the usual two-dimensional $q=2,3$ models, for which
Baillie and Coddington measure $z = 0.25(1),0.57(1)$ respectively \cite{10c}.)
Thus we see that the Wolff cluster algorithm almost eliminates
critical slowing down even on dynamical graphs.
For the graph update we use the Metropolis algorithm with the
standard ``flip'' move \cite{10d}.
As we are working with phi-cubed
graphs the detailed balance condition involves checking that the rings
at either ends of the link being flipped have no links in common.
This check also eliminates all graphs containing tadpoles or self-energies.
After each Potts model update sweep we randomly pick $NFLIP$ links one
after another and try to flip them. After testing various values of $NFLIP$
to ensure that there were enough flips to make the graph dynamical on the
time scale of the Potts model updates
we set $NFLIP=N$ for all the simulations.

\section{Results}

We measure all the standard thermodynamic quantities for the spin model:
energy $E$, specific heat $C$, magnetization $M$, susceptibility $\chi$
and correlation length $\xi$; and several properties of the graph:
acceptance rates for flips, distribution of ring lengths and
internal fractal dimension $d$.
To determine $\nu$ (actually $\nu d$) and $\beta_c$
separately, instead of from the usual three-parameter
finite-size scaling fit, for example $\xi = \xi_0 (|T-T_c|/T_c)^{-\nu}$,
we used Binder's cumulant \cite{10e}.
This is done as follows.
Binder's cumulant $U_N$ on graph with $N$ nodes is defined as
\begin{equation}
U_N = 1 - {<M^4> \over 3<M^2>^2},
\end{equation}
where $<M^4>$ is the average of the fourth power of the magnetization
and $<M^2>$ is the average of its square.
For a normal temperature-driven continuous phase transition
$U_N \rightarrow 0$ for $T > T_c$ because $M$ is
gaussian distributed about 0 at high temperature,
and $U_N \rightarrow {2 \over 3}$
for $T < T_c$  because a spontaneous magnetization $M_{sp}$
develops in the low temperature phase.
At $T = T_c$, $U_N$ has a non-trivial value which scales with $N$ according to
\begin{equation}
U_N \simeq t N^{1 \over \nu d}.
\end{equation}
Therefore the slope of $U_N$ with respect to $T$ (or $\beta$) at $T_c$
gives ${1 \over \nu d}$.
This is not much use as it stands since it involves knowledge of $T_c$,
but the $maximum$ value of the slope
scales in the same way, so we can extract $\nu d$ from
\begin{equation}
\max({dU_N \over d\beta}) \simeq N^{1 \over \nu d}.
\end{equation}
We have used this successfully to obtain the values listed in column 2
of Table 4 from the fits shown in Fig. 1.
They agree with values from KPZ scaling (Table 2).

\def\bc{\beta_c^\infty}
\begin{center}
\begin{tabular}{|c|c|c|c|} \hline
$q$& $\nu d$ & $\bc (U_N)$ & $\bc (C)$ \\[.05in]
\hline
$2$& 3.20(21)& 0.77(1) & 0.7735(12)   \\[.05in]
\hline
$3$& 2.46(12)& 0.87(1) & 0.868(1)     \\[.05in]
\hline
$4$& 2.03(12)& 0.92(1) & 0.921(1)     \\[.05in]
\hline
$10$& 1.5(4) & 1.15(1) & 1.141(1)     \\[.05in]
\hline
\end{tabular}
\end{center}
\vspace{.1in}
\centerline{Table 4: Fitted values of $\nu d$ and inverse critical temperature
$\beta_c$}
\centerline{from Binder's cumulant $U_N$ and specific heat $C$.}
\bigskip

Next, knowing $\nu d$, we use the standard finite-size scaling relation
(with $L$ replaced by $N^{1 \over d}$ since we do not know $d$ a priori)
\begin{equation}
|\beta_c^N - \beta_c^\infty| \simeq N^{-{1 \over \nu d}}
\end{equation}
to extract $\beta_c^\infty$, using $\beta_c^N$s obtained from
the position of the maximum in the slope of $U_N$ or from the
peak in the specific heat.
The latter is an order of magnitude more accurate, as can be seen from
the results in columns 3 and 4 of Table 4. The fits for $q=2$ are shown
in Fig. 2.

Lastly, we measure the other critical exponents
either from the singular behavior of the thermodynamic functions
\begin{equation}
C = B + C_0 t^{-\alpha}, \; M = M_0 t^{\beta}, \; \chi = \chi_0 t^{-\gamma},
\; \xi = \xi_0 t^{-\nu}
\label{eF1}
\end{equation}
knowing $\beta_c$,
or from the finite-size scaling relations
\begin{equation}
C = B' + C_0' N^{\alpha \over \nu d}, \; M = M_0' N^{-\beta \over \nu d}, \;
\chi = \chi_0' N^{\gamma \over \nu d}
\label{eF2}
\end{equation}
using the previously obtained value of $\nu d$.
Despite the fact that the quality of the former set of fits depends very
strongly on the precise value of $\beta_c$,
it turns out that they are better than the latter;
we have listed the values of the exponents obtained in Table 5,
columns 2-5 for the former and 6-8 for the latter.

\begin{center}
\begin{tabular}{|c|c|c|c|c|c|c|c|c|c|} \hline
$q$& $\alpha$  & $\beta$  & $\gamma$ & $\nu$ &
$\alpha/\nu d$ & $\beta/\nu d$ & $\gamma/\nu d$ & $d$ \\[.05in]
\hline
$2$& -0.98(7)  & 0.275(4) & 1.91(13) & 0.87(2) &
     0.32(1)   & 0.155(10)     & 0.79(1)        & 2.375(19) \\[.05in]
\hline
$3$& -0.48(5)  & 0.217(3) & 1.54(10) & 0.82(1) &
     0.19(1)   & 0.128(5)      & 0.81(1)        & 2.376(19) \\[.05in]
\hline
$4$& log       & 0.304(3) & 1.03(4)  & 0.65(1) &
     log       & 0.207(6)      & 0.70(1)        & 2.356(19) \\[.05in]
\hline
\end{tabular}
\end{center}
\vspace{.1in}
\centerline{Table 5: Measured values of critical exponents $\alpha, \beta,
\gamma, \nu$}
\centerline{and internal fractal dimension $d$ from $N=5000$ graphs;}
\centerline{`log' signifies that a logarithmic fit was better than a power}
\centerline{law fit implying that the corresponding exponent is $0$.}
\bigskip

\noindent
In order to constrain the first set of fits (by reducing the number of
free parameters from three to two) we fix $\beta_c$ to be the exact
value for the Ising model and the values given in column 4 of Table 4
for the $q=3,4$ Potts models.
These fits all use data from largest graphs simulated ($N=5000$).
All of the fits for the specific heat are very good because there is an
extra adjustable constant in eqs. \ref{eF1},\ref{eF2} ($B,B'$) --
we easily obtain the values predicted by KPZ for both $\alpha$ and $\alpha/\nu
d$.
The next best fits are those for $\gamma$ which again yield the expected
values.
Unfortunately the same is not true for $\beta$, the exponent governing the
vanishing of the magnetization as $T \rightarrow T_c$ from below,
where we obtain values around $1/4$ or $1/3$ rather than $1/2$.
Presumably either our graphs are not big enough to distinguish
the singularity from the ``background'' finite-size rounding or
there are large corrections to scaling (or both). We are currently
running on larger graphs to check this. One reassuring sign is that the
fitted value of $\beta$ does increase with $N$.
For the Ising model,
Catterall {\it et al} \cite{9} also had difficulty with this exponent,
estimating that $\beta = 0.25(10)$. However, Ben-Av {\it et al} \cite{8b},
who use dynamical triangulations rather than phi-cubed graphs,
manage to obtain $\beta = 0.45(10)$.
Despite the fact that plots of the scaled magnetization and susceptibility,
shown in Figs. 3 and 4, look fairly good (exhibiting the expected asymptotic
slopes of $\beta$ and $\gamma$ respectively),
we have a little difficulty with the finite-size scaling fits to both
$\beta/\nu d$ and $\gamma/\nu d$. The former comes out somewhat low
($0.155(10),0.128(5),0.207(6)$ rather than $1/6, 1/5, 1/4$ for $q=2,3,4$
respectively) and the latter somewhat high
($0.79(1),0.81(1),0.70(1)$ rather than $2/3, 3/5, 1/2$ for $q=2,3,4$
respectively). Again, comparing with the previous Ising model simulations,
we find that both Jurkiewicz {\it et al} \cite{8a}
and Catterall {\it et al} \cite{9} obtained less accurate
but consistent values $0.16(3)$ and $0.16(1)$ respectively for $\beta/\nu d$,
and $0.71(4)$ and $0.6(1)$ respectively for $\gamma/\nu d$.
{}From this it appears that our data may be becoming accurate enough to
allow examination of correction-to-scaling effects, and we shall do this when
we have
more data on larger systems.

We also fitted the exponent $\nu$ from the power law
divergence of the correlation length $\xi$ at $\beta_c$.
However, $\xi$ itself must be obtained from a fit:
the 2-point correlation function $\Gamma$ should behave as
\begin{equation}
\Gamma(r) \ \equiv \ \sum \sigma_i \sigma_{i+r} \ = \ c \ e^{- m r},
\label{eXI}
\end{equation}
where $m \equiv 1/\xi$, and the sum is over some number of measurements
made on each graph with the position of the spin $\sigma_i$ being chosen
randomly.
$r$ is the internal distance between two spins on the graph, i.e. the
fewest links between them.
As two fits are involved, the results are not particularly reliable: we
obtain $\nu = 0.87(2), 0.82(1), 0.65(1)$ for $q=2,3,4$ respectively.
As discussed above, the KPZ predictions for $\nu$ involve the
internal fractal dimension $d$ so we shall postpone further discussion
of these $\nu$ values until we estimate $d$ below.

We now turn to the properties of the graphs.
The first thing we can investigate is the acceptance rate for the
Metropolis flip move
to confirm that our graphs are really dynamical. The flip can be forbidden
either from the
graph constraints coming from the detailed balance condition or from
the energy change of the spin model, so we can decompose the flip acceptance
rate into two parts:
AL -- the fraction of randomly selected links which can be flipped
satisfying the graph constraints; and
AF -- the fraction of links satisfying the graph constraints which are
actually flipped, i.e. pass the Metropolis test using the Potts model energy
change. These quantities are shown for $q=2$ on a 2000 node graph
in Fig. 5. We immediately see that both AF and AL dip at
some $\beta < \beta_c$ but at different places.

We can also examine the distribution of ring lengths in the graph,
which is the discrete equivalent of measuring the distribution of local
Gaussian curvatures
in the continuum.
For pure quantum gravity (no spin model living on the graph)
it is possible to analytically calculate this \cite{10d}. The probability $P$
of finding a ring of length $l$ is given by
\begin{equation}
P_{N \rightarrow \infty}(l) = 16 ( {3 \over 16} )^l
{(l-2)(2l-2)! \over l!(l-1)!}
\label{ePG}
\end{equation}
which decays exponentially as $l$ increases. The minimum possible ring length
is 3.
If we plot the fraction of rings of length three (PR3) in Fig. 5
along with AF and AL we see that it has a peak very close to
the dip in AL. This is reasonable since both PR3 and AL
depend only on the graph, whereas AF depends on the Potts model.
We
plot PR3 as a function of the reduced temperature $t$ for all the $q$s
in Fig. 6,
where we see that the height of the peak increases and its position
moves closer to $t=0$ as $q$ is increased.
The $q=10$ model, for which there is no conformal field theory at all, appears
to have a peak very close to $t=0$.
Recalling that the ring length on the phi-cubed graph is equivalent to the
coordination number $q_i$ of the point $i$ at the center of the ring
on the dual triangulation, and that
the local curvature $R_i$ at this point $i$ is given as $R_i = \pi
(6-q_i)/q_i$,
we see that as $q$ (and hence $c$) increases the number of points with maximal
positive
curvature (i.e. $q_i=3$ so $R_i=\pi$) increases.
These results lend some credence to the suggestion
that the failure of KPZ scaling for $1<c<25$ may be due to
the liberation of curvature singularities at $c=1$ \cite{11}
\footnote{We investigate the interesting question of whether there is
a sudden increase in singularities as the central charge is increased
through one with multiple Potts models in \cite{14}.}.
We have no explanation as to why the flip
acceptance rates dip and the fraction of rings of length three peaks
{\it away from} the phase transition point of the Potts model for $q=2,3,4$.
We can only make one interesting observation: in simulations of
the crumpling transition of dynamically triangulated random surfaces (DTRS)
a dip in the flip acceptance rate is also found away from the transition
at $\lambda < \lambda_c$ in the crumpled phase (which corresponds to
$\beta < \beta_c$ here) \cite{10bjw}.

We now show that the distribution of ring lengths in the graphs is
determined by the central charge of the Potts model living on them.
If we plot the difference in the fraction of rings of length three at
the critical point $\beta_c$ of the Potts model from the pure gravity
fraction (eq. \ref{ePG} with $l=3$)
against the central charge of the Potts model (which we know for $q=2,3,4$),
we find a straight line with slope $0.010(1)$ which passes through the origin.
This is shown in Fig. 7, along with some results from simulations
of multiple Ising models \cite{14}. These also lie on the line,
although their central charge places them in the strong-coupling region
of Liouville theory where the KPZ results break down.
We can also plot a difference using the peak height in the fraction of rings of
length three
against the central charge to obtain another straight line with slope
$0.015(1)$ (this is also shown in Fig. 7).
However, the peak does not occur at $\beta_c$ so the correlation length is
not infinite and we cannot expect the results of conformal field theory to
apply there.
Hence it appears that if we have a model whose central charge we do not know
we can look up the value of PR3 at the phase transition (or its peak value) on
the y-axis of
Fig. 7 and read off its ``effective central charge'' from the
x-axis.
If this relation holds in general
it would provide a viable method of obtaining $c$ for any model coupled to
quantum gravity either on a random graph or on a DTRS. Interestingly
the $q=10$ Potts model still lies on the line
despite the absence of a corresponding conformal
field theory, giving an ``effective central charge'' consistent with $1$.

To complete our discussion of the graph properties,
we look at their internal fractal dimension $d$.
We use the most naive definition of distance (the fewest links between two
nodes)
so we are considering the ``mathematical geometry'' rather than the ``physical
geometry'' in the terminology of \cite{12a}.
The values obtained for graphs with $5000$ nodes are listed in the last
column of Table 5. These were measured at the critical point of the
Potts model but the same results (within statistical error bars) were
obtained for all $\beta$. Moreover the same value of $d$ (within errors)
was obtained for each $q$. The values do, however, depend on $N$ and
if we extrapolate to $N=\infty$ for the Ising model we obtain $d=2.78(4)$.
On very large graphs of around 100,000 nodes with no matter
(pure two-dimensional quantum gravity)
Agishtein and Migdal
\cite{12b} found that the relation between the area $V(r)$ and radius $r$ of
a circle using the mathematical geometry was of the form
\begin{equation}
\log V = a + b \log ( r ) + c ( \log ( r ) )^2
\end{equation}
so there was no fractal dimension at all. It is possible that a similar effect
may be found
when simulations such as ours incorporating matter are carried out on
graphs some orders of
magnitude bigger \footnote{It is a much more demanding problem to generate huge
graphs with matter
as it is no longer possible to use graph enumeration formulae to generate them
recursively
as was done for pure quantum gravity.}.
If this is the case it is difficult to understand how critical behavior, which
assumes some kind
of scaling and hence fractal dimension, can appear at all. Nonetheless, the
exact solution
of the Ising model, the KPZ results and the simulations in this and other
papers appear to show
that phase transitions are taking place and critical exponents can be defined.
A possible
resolution of this problem is that the spin degrees of freedom are actually
sensitive
to the physical geometry which is less singular than the mathematical geometry.
A sign that this is indeed the case may be found in the fact that
Agishtein and Migdal obtain a value of $2.7$ for the internal fractal dimension
of their pure quantum gravity graphs
using the physical geometry \cite{12a}, which is surprisingly close to our
$2.78(4)$.
Ignoring these qualms about the existence of the fractal dimension,
we resume our discussion of the $\nu$ values obtained above from the
correlation length.
Taking our value of $d=2.78$ we estimate $\nu d = 2.4, 2.3, 1.8$,
whereas KPZ scaling predicts $3,2.5,2$ (for $q=2,3,4$ respectively).
There is obviously some discrepancy but the numbers are fairly close,
implying that our analysis is consistent.

Finally, we briefly discuss our results for the $q=10$ Potts model.
On a fixed graph the $q >4$ Potts models display first order transitions, so
there
is no corresponding conformal field theory. The fact that the $q=4$ Potts model
lies at the boundary of the region where the KPZ formula applies ($c=1$)
suggests that something similar might happen when $q>4$ Potts models are
coupled to quantum gravity.
By examining the behavior of Binder's cumulant \cite{10e} it is clear
that the $q=10$ Potts model retains its first order phase transition on
a dynamical graph.
As shown in Fig. 8, $U_N$ has a minimum (the position of which
$\rightarrow \beta_c$ as $N \rightarrow \infty$) as expected for a first order
transition, and tends to $1/2$ (rather than to $0$ which is the case
for a second order transition) as $\beta \rightarrow 0$.
{}From finite-size scaling we expect that the peaks in the specific heat and
susceptibility grow as $L^d$, i.e. $N$, for first order phase transitions.
If we fit $\max(\chi)$ versus $N$ then we obtain an exponent $0.93(1)$.
(We can also fit to $\max(C)$ but as before the adjustable constant $B$
renders the fit insignificant.)

\section{Conclusions}

To summarize, we have verified numerically that
the critical exponents for the $q=2,3,4$ Potts models on dynamical phi-cubed
graphs
(i.e. coupled to two-dimensional quantum gravity) are in reasonable agreement
with those predicted by KPZ,
and that the $q=10$ Potts model appears to have a first order transition. We
have also found
some interesting behavior in the graphs themselves, namely that there is a peak
in the ring (curvature)
distribution which approaches $\beta_c$ from below as $c \rightarrow 1$,
and that there is a linear relation
between the probability of rings of length three and $c$.
{}From the algorithmic point of view, our measurement of the dynamical critical
exponent $z$
reveals that the Wolff algorithm is effective in alleviating critical slowing
down
for Potts models
on {\it dynamical} graphs as well as on fixed lattices.
In companion papers we explore the fractal properties of
the spin-clusters that we use in our Wolff algorithm \cite{13},
comparing them with those on fixed graphs, and
use multiple copies of the Potts models to explore the internal geometry of the
graph in
the strong coupling region \cite{14}.

\bigskip
\bigskip
\centerline{\bf Acknowledgements}
\bigskip

This work was supported in part by NATO collaborative research grant CRG910091.
CFB is supported by DOE under contract DE-AC02-86ER40253 and by AFOSR Grant
AFOSR-89-0422.
We would like to thank M.E. Agishtein and A.A. Migdal
for providing us with initial graphs generated by
their two-dimensional quantum gravity code, and S. Catterall and
A. Krzywicki for useful discussions.

\vfill
\eject

\vfill
\eject
\centerline{\bf Figure Captions}
\begin{description}
\item{Fig. 1.}
Fit to maximum slope of derivative of Binder's cumulant versus $N$ to extract
$\nu d$.
\item{Fig. 2.}
Extrapolation of $\beta_c^N$ from Binder's cumulant and specific heat
to estimate $\beta_c^\infty$ for Ising model.
\item{Fig. 3.}
Finite-size scaling plot of $M$ for (inverse temperature) $\beta < \beta_c$,
with
expected asymptotic slope of (exponent) $\beta = 0.5$ for all models shown as
line.
\item{Fig. 4.}
Finite-size scaling plot of $\chi$ for $\beta < \beta_c$, with
expected asymptotic slopes of $\gamma = 2,1.5,1$ for
the $q=2,3,4$ Potts models respectively shown as lines.
\item{Fig. 5.}
AF, AL and PR3 for Ising model on graph with $N=2000$;
the y-scale applies to AF only, AL and PR3 have been scaled
appropriately to fit on plot; $\beta_c$ is indicated by vertical line.
\item{Fig. 6.}
Probabilities of rings of length three PR3 as function of reduced temperature
$t$ for all $q$.
\item{Fig. 7.}
Difference in PR3 at $\beta_c$ of Potts model, and at its peak,
from the pure quantum gravity value versus the central charge $c$,
for multiple Ising models as well as for single Potts models.
\item{Fig. 8.}
Binder's cumulant for $q=10$ Potts model (errors bars
omitted for clarity); $\beta_c$ is indicated by vertical line.
\end{description}
\end{document}